%%%%%%% version 11/4/2004 %%%%%%%%%%%%%%%%%
\input harvmac
\input epsf
\input amssym

\noblackbox

\def\coeff#1#2{\relax{\textstyle {#1 \over #2}}\displaystyle}

\def\whQ{\widehat Q}

\def\bfone{\relax{\rm 1\kern-.35em 1}}
\def\IC{\relax\,\hbox{$\inbar\kern-.3em{\rm C}$}}
\def\ID{\relax{\rm I\kern-.18em D}}
\def\IF{\relax{\rm I\kern-.18em F}}
\def\IH{\relax{\rm I\kern-.18em H}}
\def\II{\relax{\rm I\kern-.17em I}}
\def\IN{\relax{\rm I\kern-.18em N}}
\def\IP{\relax{\rm I\kern-.18em P}}
\def\IQ{\relax\,\hbox{$\inbar\kern-.3em{\rm Q}$}}

\def\IR{\relax{\rm I\kern-.18em R}}
\font\cmss=cmss10 \font\cmsss=cmss10 at 7pt
\def\ZZ{\relax\ifmmode\mathchoice
{\hbox{\cmss Z\kern-.4em Z}}{\hbox{\cmss Z\kern-.4em Z}}
{\lower.9pt\hbox{\cmsss Z\kern-.4em Z}}
{\lower1.2pt\hbox{\cmsss Z\kern-.4em Z}}\else{\cmss Z\kern-.4em
Z}\fi}

\def\ZZ{\Bbb{Z}} 
\def\IC{\Bbb{C}}
\def\ID{\Bbb{D}}
\def\IF{\Bbb{F}}
\def\IH{\Bbb{H}}
\def\II{\Bbb{I}}
\def\IN{\Bbb{N}}
\def\IP{\Bbb{P}}
\def\IQ{\Bbb{Q}}
\def\IR{\Bbb{R}}
%
%
 %%%%%%%%%%%%%  References %%%%%%%%%%%%%
 % 
 %\LuninFV
\lref\LuninFV{
O.~Lunin and S.~D.~Mathur,
``Metric of the multiply wound rotating string,''
Nucl.\ Phys.\ B {\bf 610}, 49 (2001)
[arXiv:hep-th/0105136].
%%CITATION = HEP-TH 0105136;%%
}
%
%\MateosQS
\lref\MateosQS{
D.~Mateos and P.~K.~Townsend,
``Supertubes,''
Phys.\ Rev.\ Lett.\  {\bf 87}, 011602 (2001)
[arXiv:hep-th/0103030].
%%CITATION = HEP-TH 0103030;%%
}
%\LuninJY
\lref\LuninJY{
O.~Lunin and S.~D.~Mathur,
``AdS/CFT duality and the black hole information paradox,''
Nucl.\ Phys.\ B {\bf 623}, 342 (2002)
[arXiv:hep-th/0109154].
%%CITATION = HEP-TH 0109154;%%
}
%
%\LuninQF
\lref\LuninQF{
O.~Lunin and S.~D.~Mathur,
``Statistical interpretation of Bekenstein entropy for systems with a
stretched horizon,''
Phys.\ Rev.\ Lett.\  {\bf 88}, 211303 (2002)
[arXiv:hep-th/0202072].
%%CITATION = HEP-TH 0202072;%%
}
%
%\LuninBJ
\lref\LuninBJ{
O.~Lunin, S.~D.~Mathur and A.~Saxena,
``What is the gravity dual of a chiral primary?,''
Nucl.\ Phys.\ B {\bf 655}, 185 (2003)
[arXiv:hep-th/0211292].
%%CITATION = HEP-TH 0211292;%%
}
%
%\LuninUU
\lref\LuninUU{
O.~Lunin,
``Adding momentum to D1-D5 system,''
JHEP {\bf 0404}, 054 (2004)
[arXiv:hep-th/0404006].
%%CITATION = HEP-TH 0404006;%%
}
%
%%%%%%%%%%%%%%%%%%%%%%%%%%%%%%
%
%\MathurHJ
\lref\MathurHJ{
S.~D.~Mathur, A.~Saxena and Y.~K.~Srivastava,
``Constructing 'hair' for the three charge hole,''
Nucl.\ Phys.\ B {\bf 680}, 415 (2004)
[arXiv:hep-th/0311092].
%%CITATION = HEP-TH 0311092;%%
}
%
%\MathurSV
\lref\MathurSV{
S.~D.~Mathur,
``Where are the states of a black hole?,''
arXiv:hep-th/0401115.
%%CITATION = HEP-TH 0401115;%%
}
%
%\GiustoID
\lref\GiustoID{
S.~Giusto, S.~D.~Mathur and A.~Saxena,
``Dual geometries for a set of 3-charge microstates,''
arXiv:hep-th/0405017.
%%CITATION = HEP-TH 0405017;%%
}
%
%\GiustoIP
\lref\GiustoIP{
S.~Giusto, S.~D.~Mathur and A.~Saxena,
``3-charge geometries and their CFT duals,''
arXiv:hep-th/0406103.
%%CITATION = HEP-TH 0406103;%%
}
%
%\GiustoKJ
\lref\GiustoKJ{
S.~Giusto and S.~D.~Mathur,
``Geometry of D1-D5-P bound states,''
arXiv:hep-th/0409067.
%%CITATION = HEP-TH 0409067;%%
}
%%\BenaWT
\lref\BenaWT{
I.~Bena and P.~Kraus,
``Three charge supertubes and black hole hair,''
Phys.\ Rev.\ D {\bf 70}, 046003 (2004)
[arXiv:hep-th/0402144].
%%CITATION = HEP-TH 0402144;%%
}
%\BenaDE
\lref\BenaDE{
I.~Bena and N.~P.~Warner,
``One ring to rule them all ... and in the darkness bind them?,''
arXiv:hep-th/0408106.
%%CITATION = HEP-TH 0408106;%%
}
%\BenaTK
\lref\BenaTK{
I.~Bena and P.~Kraus,
``Microscopic description of black rings in AdS/CFT,''
arXiv:hep-th/0408186.
%%CITATION = HEP-TH 0408186;%%
}
%
%\BenaWV
\lref\BenaWV{
I.~Bena,
``Splitting hairs of the three charge black hole,''
arXiv:hep-th/0404073.
%%CITATION = HEP-TH 0404073;%%
}
%
%\BenaJW
\lref\BenaJW{
I.~Bena and N.~P.~Warner,
``A harmonic family of dielectric flow solutions with maximal supersymmetry,''
arXiv:hep-th/0406145.
%%CITATION = HEP-TH 0406145;%%
}
%
%\LinNB
\lref\LinNB{
H.~Lin, O.~Lunin and J.~Maldacena,
``Bubbling AdS space and 1/2 BPS geometries,''
JHEP {\bf 0410}, 025 (2004)
[arXiv:hep-th/0409174].
%%CITATION = HEP-TH 0409174;%%
}
%
%\ElvangDS
\lref\ElvangDS{
H.~Elvang, R.~Emparan, D.~Mateos and H.~S.~Reall,
``Supersymmetric black rings and three-charge supertubes,''
arXiv:hep-th/0408120.
%%CITATION = HEP-TH 0408120;%%
}
%
%\ElvangRT
\lref\ElvangRT{
H.~Elvang, R.~Emparan, D.~Mateos and H.~S.~Reall,
``A supersymmetric black ring,''
arXiv:hep-th/0407065.
%%CITATION = HEP-TH 0407065;%%
}
%
%\ElvangMJ
\lref\ElvangMJ{
H.~Elvang and R.~Emparan,
``Black rings, supertubes, and a stringy resolution of black hole
non-uniqueness,''
JHEP {\bf 0311}, 035 (2003)
[arXiv:hep-th/0310008].
%%CITATION = HEP-TH 0310008;%%
}
%
%\ElvangYY
\lref\ElvangYY{
H.~Elvang,
``A charged rotating black ring,''
Phys.\ Rev.\ D {\bf 68}, 124016 (2003)
[arXiv:hep-th/0305247].
%%CITATION = HEP-TH 0305247;%%
}
%
%\EmparanWN
\lref\EmparanWN{
R.~Emparan and H.~S.~Reall,
``A rotating black ring in five dimensions,''
Phys.\ Rev.\ Lett.\  {\bf 88}, 101101 (2002)
[arXiv:hep-th/0110260].
%%CITATION = HEP-TH 0110260;%%
}
%
%
%\EmparanWY
\lref\EmparanWY{
R.~Emparan,
``Rotating circular strings, and infinite non-uniqueness of black rings,''
JHEP {\bf 0403}, 064 (2004)
[arXiv:hep-th/0402149].
%%CITATION = HEP-TH 0402149;%%
}
%\GauntlettQY
\lref\GauntlettQY{
J.~P.~Gauntlett and J.~B.~Gutowski,
``General concentric black rings,''
arXiv:hep-th/0408122.
%%CITATION = HEP-TH 0408122;%%
}
%
%\GauntlettWH
\lref\GauntlettWH{
J.~P.~Gauntlett and J.~B.~Gutowski,
``Concentric black rings,''
arXiv:hep-th/0408010.
%%CITATION = HEP-TH 0408010;%%
}
%
%\SenIN
\lref\SenIN{
A.~Sen,
``Extremal black holes and elementary string states,''
Mod.\ Phys.\ Lett.\ A {\bf 10}, 2081 (1995)
[arXiv:hep-th/9504147].
%%CITATION = HEP-TH 9504147;%%
}
%
%\StromingerSH
\lref\StromingerSH{
A.~Strominger and C.~Vafa,
``Microscopic Origin of the Bekenstein-Hawking Entropy,''
Phys.\ Lett.\ B {\bf 379}, 99 (1996)
[arXiv:hep-th/9601029].
%%CITATION = HEP-TH 9601029;%%
}
%
%\GauntlettNW
\lref\GauntlettNW{ 
J.~P.~Gauntlett, J.~B.~Gutowski, C.~M.~Hull, S.~Pakis and H.~S.~Reall,
``All supersymmetric solutions of minimal supergravity in five dimensions,''
Class.\ Quant.\ Grav.\  {\bf 20}, 4587 (2003)
[arXiv:hep-th/0209114].
%%CITATION = HEP-TH 0209114;%%
%
}
%\GutowskiYV
\lref\GutowskiYV{
J.~B.~Gutowski and H.~S.~Reall,
``General supersymmetric AdS(5) black holes,''
JHEP {\bf 0404}, 048 (2004)
[arXiv:hep-th/0401129].
%%CITATION = HEP-TH 0401129;%%
}
\lref\gary{
 G.~T.~Horowitz and H.~S.~Reall,
  %``How hairy can a black ring be?,''
  Class.\ Quant.\ Grav.\  {\bf 22}, 1289 (2005)
  [arXiv:hep-th/0411268].
  %%CITATION = HEP-TH 0411268;%%
}
%
%\PalmerGU
\lref\PalmerGU{
B.~C.~Palmer and D.~Marolf,
``Counting supertubes,''
JHEP {\bf 0406}, 028 (2004)
[arXiv:hep-th/0403025].
%%CITATION = HEP-TH 0403025;%%
}
%\BreckenridgeIS
\lref\BreckenridgeIS{
J.~C.~Breckenridge, R.~C.~Myers, A.~W.~Peet and C.~Vafa,
``D-branes and spinning black holes,''
Phys.\ Lett.\ B {\bf 391}, 93 (1997)
[arXiv:hep-th/9602065].
%%CITATION = HEP-TH 9602065;%%
}
%\TseytlinAS
\lref\TseytlinAS{
A.~A.~Tseytlin,
``Extreme dyonic black holes in string theory,''
Mod.\ Phys.\ Lett.\ A {\bf 11}, 689 (1996)
[arXiv:hep-th/9601177].
%%CITATION = HEP-TH 9601177;%%
}
%
%\BakKZ
\lref\BakKZ{
D.~Bak, Y.~Hyakutake, S.~Kim and N.~Ohta,
``A geometric look on the microstates of supertubes,''
arXiv:hep-th/0407253.
%%CITATION = HEP-TH 0407253;%%
}
%
%\BakRJ
\lref\BakRJ{
D.~Bak, Y.~Hyakutake and N.~Ohta,
``Phase moduli space of supertubes,''
Nucl.\ Phys.\ B {\bf 696}, 251 (2004)
[arXiv:hep-th/0404104].
%%CITATION = HEP-TH 0404104;%%
}
%
%\GowdigereJF
\lref\GowdigereJF{
C.~N.~Gowdigere, D.~Nemeschansky and N.~P.~Warner,
``Supersymmetric solutions with fluxes from algebraic Killing spinors,''
Adv.\ Theor.\ Math.\ Phys.\  {\bf 7}, 787 (2004)
[arXiv:hep-th/0306097].
%%CITATION = HEP-TH 0306097;%%
}
%
%\PilchJG
\lref\PilchJG{
K.~Pilch and N.~P.~Warner,
``Generalizing the N = 2 supersymmetric RG flow solution of IIB supergravity,''
Nucl.\ Phys.\ B {\bf 675}, 99 (2003)
[arXiv:hep-th/0306098].
%%CITATION = HEP-TH 0306098;%%
}
%

%%%%%%%%%%%%% End References %%%%%%%%%%%%%
%%%%%%%%%%%%%%%%%%%%%%%%%%%%%%%%%%%%%%%%

%%%%%%%%%%%%%%%   Title Page  %%%%%%%%%%%%%

\Title{
\vbox{
\hbox{\tt hep-th/0411072}
\hbox{\tt  UCLA/04/TEP/45}
}}
{\vbox{\vskip -1.0cm
\centerline{\hbox{Black Rings with Varying Charge Density}}}}
\vskip -.3cm
\centerline{Iosif~Bena${}^{(1)}$,  Chih-Wei Wang${}^{(2)}$ and 
Nicholas P.\ Warner${}^{(2)}$}   
\bigskip
\centerline{{${}^{(1)}$\it Department of Physics and Astronomy}}
\centerline{{\it University of California}}
\centerline{{\it Los Angeles, CA  90095, USA}}
\medskip
\centerline{{${}^{(2)}$\it Department of Physics and Astronomy}}
\centerline{{\it University of Southern California}}
\centerline{{\it Los Angeles, CA 90089-0484, USA}}
\medskip

\bigskip
\bigskip
We find the general five-dimensional, supersymmetric black ring solutions in $M$-theory
based upon a circular ring, but with arbitrary, fluctuating charge distributions 
around the ring.   The solutions have three arbitrary charge distribution functions,
but their asymptotic charges and angular momenta only depend upon
the total charges on the ring. The arbitrary density fluctuations thus represent ``hair.''
By varying the charge distributions one can continuously change  the entropy of 
these black rings; to our knowledge this is the first solution in which the entropy 
depends on classical moduli. We also show that there is a family of solutions, with
two arbitrary functions,  for which the horizon remains rotationally  invariant, and yet the 
complete  solution breaks rotational symmetry.    
If the horizon area is set to zero then one obtains families of 
supertube solutions.  We find that our general solutions are governed by 
three harmonic functions that may be thought of as classical excitations of a string.  
The horizon area provides a natural Lorentz metric on these excitations, and  the 
constancy of the  rotational invariance of the horizon imposes a set of Virasoro constraints.

\vskip .3in
%\draft
\Date{\sl {November, 2004}}

\vfill\eject

%%%%%%%%%%%%%%%%%%%%%%%%%%%%%%%%%%
\newsec{Introduction}
%%%%%%%%%%%%%%%%%%%%%%%%%%%%%%%%%%

There are quite a number of reasons to  study
and classify  three-charge, BPS black ring and supertube solutions.
For relativists  it is interesting to understand the large variety 
of such solutions 
\refs{\BenaWV\ElvangRT\BenaDE\ElvangDS\GauntlettWH{--}\GauntlettQY}
and the large extent to which they violate black hole uniqueness. 
Another major driving force is the recent proposal by Mathur and collaborators
that supertubes can be thought of as individual 
black-hole microstates and that black holes should be thought of as
``statistical ensembles'' of regular horizon-less microstate geometries 
\refs{\LuninFV\LuninJY\LuninQF\LuninBJ\MathurHJ\MathurSV\GiustoID
\GiustoIP{--}\GiustoKJ}.

Although string theory indicates the existence of large families of three-charge
BPS solutions \BenaWT, finding these solutions has proven to be quite 
hard given the complexity of the underlying equations. The
explicitly-known,  three-charge BPS solutions are: the BMPV black hole 
\refs{\BreckenridgeIS,\TseytlinAS}, 
supersymmetric three-charge supertubes, black rings 
with horizon topology $S^1 \times S^2$ in minimal supergravity \ElvangRT, or 
eleven-dimensional supergravity \refs{\BenaDE,\ElvangDS,\GauntlettQY,\BenaWV}, 
superpositions of 
black rings \refs{\GauntlettWH,\GauntlettQY}, as well as solutions 
obtained by other methods \refs{\MathurHJ, \LuninUU,\MathurSV\GiustoID
\GiustoIP{--}\GiustoKJ}.  All the explicit solution are $U(1) \times U(1)$ invariant 
\foot{Some implicit solutions with only one $U(1)$ invariance have also been constructed
 \refs{\GauntlettWH,\GauntlettQY}}.  
However, if one is to 
use these solutions to explain black hole entropy \MathurSV\ or to find the extent of
the violation of black hole uniqueness, it is important to find three-charge BPS supertubes, 
black rings and other solutions that do not have such a large symmetry.

By analyzing the Killing spinors of three-charge solutions using the Killing spinor 
methods developed in \refs{\GowdigereJF,\PilchJG},  it was shown in
\BenaDE\ that the general problem of finding three-charge solutions
that preserve the same supersymmetries as the three-charge black hole
can be reduced to solving {\it linear} equations of ordinary, Euclidean,
four-dimensional  electromagnetism.  While these equations appear,
at first sight, to be non-linear\foot{These equations were also 
found in work that classifies five-dimensional supergravity solutions 
\refs{\GauntlettNW,\GutowskiYV}.}, 
it was also pointed out in \BenaDE\ that if these equations
are solved in the proper sequence, then the non-linearities only appear
in source terms, meaning that the problem is actually linear. 
To find solutions with a flat base, one first choses an arbitrary profile 
shape, $\vec \mu(\psi) \in \IR^4$, $0 \le \psi \le 2 \pi$,  for the ring, 
which determines the fluxes sourced by the dipole branes. One then is free to 
choose three arbitrary charge density functions, $\rho_i(\psi)$, around the ring, and 
determine  the harmonic functions sourced by these charges, which are then used to find 
the angular momentum and complete the solution. There are thus seven 
{\it arbitrary functions} in the general solution
of \BenaDE.  While this general solution with seven arbitrary functions is not
explicitly exhibited in \BenaDE, its existence is guaranteed by the existence
of solutions in Euclidean  electromagnetism for a given source distribution.

Even though a solution to the equations of electromagnetism may exist, the
corresponding solution in $M$-theory is required to satisfy some more
stringent physical conditions:  Most particularly, there should be no closed
time-like curves (CTCs).  In \BenaDE\ it was argued that any supertube solution, when
approached sufficiently closely, should look like the straight supertube of
\BenaWV, and hence be regular, and free of CTC's.  However, this observation
really only guarantees that there are not ``local'' CTC's around the ring, and also does not 
guarantee the absence of Dirac strings. It is therefore important to construct these
solutions and examine their features explicitly.  

In this paper we construct new, asymptotically flat, black ring solutions given 
by three of the seven arbitrary functions that govern 
general solutions with a flat base. The black rings and three-charge supertubes 
have circular dipole profiles, but have arbitrary charge
densities.  We will see that the charge densities feed back into the geometrical
shape of the ring, causing its physical radius of curvature to vary. We also verify that the 
solutions have no  Dirac strings and no CTC's  near the ring. Although we have 
not analyzed whether the metric at the horizon is smooth, there is a family of our solutions, 
parametrized by two arbitrary functions, that  has the same near-horizon geometry as the 
$U(1) \times U(1)$ invariant black ring, and intuitively one should expect it to be completely 
smooth at the horizon\foot{The continuity issues 
at the horizon are rather subtle, and are being investigated
by Horowitz and Reall \gary.}.

Our solutions describe families of black ring solutions that have three, 
freely choosable charge densities around the ring, but whose asymptotic charges are 
the same as those of the $U(1) \times U(1) $  black ring with the same total charges.  
The entropy of these rings will turn out to be the integral of a certain 
functional of these functions,  and can thus be freely varied by changing the 
charge densities. This freedom  persists in the two parameter family that is 
$U(1) \times U(1)$ invariant near the horizon.  Hence, our solutions are the first example of a 
black hole whose entropy depends on  continuous parameters --
the charge densities along the ring. 

To find the supertube solutions one wants to set the horizon area to zero, but
we consider the more general problem of setting the horizon area equal to a constant.
This is interesting because it results in families of solutions that break the 
$U(1)$ symmetry around the ring axis, but that have ``round'' horizons, that is, the
horizon {\it does} have rotational invariance along the ring.  Setting the
horizon area to a constant imposes one functional condition, and we have three
arbitrary functions in the charge densities, and so there are generally going to be 
families of solutions with constant horizons parametrized by two arbitrary functions.

We also find a rather simpler, and physically very interesting way to characterize our 
solutions.  By changing coordinates, one may think of our solutions as 
being governed by three harmonic functions on a semi-infinite cylinder.  The
charge densities source the functions at the bottom of the cylinder, and the 
solutions are required to vanish at infinity.  One may thus think of the solutions
as some form of bosonic string excitations theory.   The horizon radius of curvature 
provides a simple Lorentz metric on this bosonic string, and requiring that the radius of 
curvature  be constant (a ``round'' horizon)  amounts to imposing a set of Virasoro 
constraints. While these statements are classical, it does suggest that the charge excitations
of the black ring might be quantized in terms of such a bosonic string.   We will
defer a more detailed investigation of these issues to a subsequent paper.

 In the next section we will summarize the results of  \BenaDE\ concerning the
 system of equations that govern supertubes and black rings.  In section 3 we present the 
 solutions with varying charge densities, while section 4 discusses the
 solutions with round horizons, and contains a simple example.
 In section 5 we recast our results in terms of a classical bosonic string, 
 and we make some final comments  in section 6.

 %%%%%%%%%%%%%%%%%%%%%%%%%%%%%%%%%%
\newsec{The black-ring equations}
%%%%%%%%%%%%%%%%%%%%%%%%%%%%%%%%%%

The simplest way to describe the three-charge black ring is in terms of
three sets of M2 branes.  We take the M2 branes to lie in the $123$, $145$ and
$167$ directions, so that they wrap tori in the spatial directions.  
The non-compact space-time is thus in the $1\,8\,9\,10\,11$ directions.
The metric Ansatz is:
\eqn\metansatz{\eqalign{
e^1  ~=~ & e^{-2{A_1} - 2{A_2} - 2{A_3}} \big(\, {dx}^{1} + \vec k \cdot d\vec y\,  \big),  \cr
e^2~=~  & e^{-2{A_1} + {A_2} + {A_3}} {dx}^{2}\,, \qquad
e^3~=~    e^{-2{A_1} + {A_2} + {A_3}} {dx}^{3}\,,\cr
e^4 ~=~ &  e^{{A_1} - 2{A_2} + {A_3}}{dx}^{4}\,, \qquad
e^5 ~=~   e^{{A_1} - 2{A_2} + {A_3}}{dx}^{5}\,, \cr
e^6 ~=~ &  e^{{A_1} + {A_2} - 2{A_3}}{dx}^{6}\,, \qquad
e^7 ~=~  e^{{A_1} + {A_2} - 2{A_3}}{dx}^{7}\,,\cr
e^{7+j} ~=~ & e^{{A_1} + {A_2} +{A_3}}{dy^j}\,, \quad j=1,\dots,4\,. }}
The solution contains both $M2$ branes and $M5$ branes.  The $M5$-branes
wrap four-tori in the $4567$, $2367$ and $2345$ directions respectively, and
their last spatial dimension will define the ring profile\foot{As was pointed out in 
\BenaDE, the three $M5$ branes  could have separate profiles, but we will not 
consider this possibility here since such independent profiles are almost certainly 
not bound states.}, $\vec y = \vec \mu(\psi) \in \IR^4$, $0 \le \psi \le 2 \pi$.
The electric charges of the $M2$ branes are completely fixed in terms
of the metric functions via the usual ``zero-force'' BPS conditions.  However,
the presence of the $M5$ branes is reflected in three independent Maxwell
fields, $\vec a_{(i)} $, in the $\IR^4$:
\eqn\Cansatz{\eqalign{
C^{(3)} ~=~ & - e^1 \wedge e^2 \wedge e^3 - e^1 \wedge e^4 \wedge e^5 -
e^1 \wedge e^6 \wedge e^7 + \cr
& ~+~  2 \,(\vec a_{(1)} \cdot d \vec y)  \wedge dx^2 \wedge dx^3 + 2 \,(\vec a_{(2)} 
\cdot  d \vec y )  \wedge dx^4 \wedge dx^5 +  2 \,(\vec a_{(3)} \cdot d \vec y)  
\wedge dx^6 \wedge dx^7 \,.}}

It is convenient to introduce the functions, $Z_j$, and the Maxwell
field strengths 
%$G_{(0)}$ and 
$G_{(j)}$, defined by:
\eqn\ZGdefn{Z_j~\equiv~ e^{6 A_j}\,, 
%\qquad G_{(0)} ~\equiv~ d \, k \,, 
\qquad G_{(j)}~\equiv~  d (a_{(j)})\,,\qquad  j =1,2,3\,.}
The equations that define the ${1 \over 8}$-BPS rings are then:
\eqn\fundeqnsa{G_{(i)}~=~  * G_{(i)}\,, }
\eqn\fundeqnsb{ d * d Z_i ~=~  2\, \sum_{j,k}\, |\epsilon^{ijk} |  \,G_{(j)} \wedge G_{(k)}\,,}
\eqn\fundeqnsc{
%G_{(0)} + * G_{(0)}  
dk+ *dk ~=~  2\, G_{(1)} Z_1 +2 \, G_{(2)} Z_2 + 2 \, G_{(3)} 
Z_3\,,}
where $*$ denotes the dual on $\IR^4$ and $k$ is the angular momentum vector 
appearing in $e^1$ in \metansatz.  If one solves this system in the order presented
here, then the system is linear.    

The general solution for a ring profile $\vec y= \vec \mu(\psi)$ can then be written in
terms of the usual Green functions, as follows.  First one computes:
\eqn\Maxsolna{\vec b_{(j)}(\vec y) ~\equiv~ {q_j \over 2 \,\pi}\, \int_0^{2 \pi}\, 
{  {\vec \mu_j}{}' (\psi) \over  | \vec y - \vec \mu_j (\psi)|^2 }   \,
d \psi\,,   }
and then sets: 
\eqn\Maxsolnb{G_{(j)} ~=~ (1+*) \,(d \, (\vec b_{(j)}\cdot d \vec y) \,)\,. }
One then uses this to obtain $\vec a_{(j)}$.   Note that the charge density
in the integrand of \Maxsolna\ is constant, reflecting the fact
that the number of $M5$ branes is constant   along the
profile.  Having solved for $G_{(j)} $, one then gets  the $Z_i$ from:
\eqn\Zeqn{
Z_i(\vec y) = 1+ \int{ \rho_i(\vec z) + 2 \, \sum_{j,k}\, |\epsilon^{ijk}|  \,
* (G_j \wedge G_k)( \vec z ) \over (\vec y- \vec z)^2 }\, d^4 z \,,}
where the constant of integration has been set to one so that
the metric has the proper asymptotics at infinity.  The functions,  
$\rho_i(\vec z)$, are freely choosable, but for a black ring or 
three-charge supertube we 
require the density functions to be supported on the ring.
Finally, one gets the angular momentum vector, $\vec k$, from
solving:
\eqn\keqn{* d * d k = * \, 2\, [(dZ_1) \wedge  G_1  + (d Z_2) \wedge  G_2  + 
 (d Z_3) \wedge G_3 ]  ~\equiv~ J \,,}
 %\
via the Green function:
\eqn\ksoln{\vec k (\vec y) ~\equiv~ \int_{\IR^4}{  \vec J(\vec z)  \over 
| \vec y - \vec z|^2 } \,d^4 z \,.}
In principle one can add a homogeneous component to the solution,
\ksoln, that is, a vector field, $\vec k_0$, for which $dk_0 + *dk_0 = 0$
outside the ring.  Such a solution would be sourced by an ``angular
momentum density,'' $\sigma(\psi)$, around the ring.  As was noted
in \BenaDE, this homogeneous solution, and the associated
density, $\sigma(\psi)$, must usually be set to zero in order to avoid CTC's.

%%%%%%%%%%%%%%%%%%%%%%%%%%%%%%%%%%
\newsec{The solutions with varying charge densities}
%%%%%%%%%%%%%%%%%%%%%%%%%%%%%%%%%%

We now seek ring solutions with varying charge densities. We will
start with a ``round'' $M5$-brane distribution, exactly as in \BenaDE,
but we will the introduce general distributions of $M2$-brane charge
around the ring.  We consider a ring of $M5$ branes located 
at   $r=0$ and $z=R$, in the coordinates $(z,\psi)$ and $(r,\phi)$ in 
which the metric on $\IR^4$ is
\eqn\firstRmet
{d\vec y \cdot d \vec y ~=~ (dz^2 ~+~ z^2 \, d \psi^2) ~+~
(dr^2 ~+~ r^2 \, d \phi^2)  \,.}
To solve the Laplace equation and to simplify the form of the $G_i$ it is simpler
to use a better-adapted set of coordinates \ElvangRT:
\eqn\xycoords{
x ~=~  -{ z^2+r^2 - R^2 \over  \sqrt{((z-R)^2 + r^2)( (z+R)^2 + r^2 )}} \,, \qquad
y ~=~ - { z^2+r^2 + R^2 \over  \sqrt{((z-R)^2 + r^2)( (z+R)^2 + r^2 )}} \,,}
for which one has $ -1 \le x \le 1$, $-\infty < y \le -1$, and the ring is located at
$y = -\infty$.  In these coordinates, the metric on $\IR^4$ becomes:
\eqn\newRmet{ ds^2_{\IR^4}= {R^2 \over (x-y)^2}\, \bigg( {dy^2 \over y^2 -1} 
+ (y^2-1)\, d \psi^2  +{dx^2 \over 1-x^2} + (1-x^2) \, d \phi^2  \bigg)\,.}

\subsec{The new solutions}

Since the shapes of the dipole branes are the same as in references 
\refs{\BenaDE,\ElvangDS,\GauntlettQY},  the fields $\vec a_{(j)}$ are the same, and hence:
\eqn\magG{
G_{(j)} ~=~  q_j \, (d x \wedge d \phi ~-~ d y \wedge d \psi) \,.}
The new element in the solution here is to allow a general charge density
around the ring.  This means that   $Z_i$ contains a term of the form:
\eqn\lambdaform{\lambda_i (z,r,\psi) ~=~ \int_0^{2\,\pi} \, 
{ \rho_i(\chi) \, R \over r^2 +  z^2 +  R^2 -  2\, z\, R\, \cos(\psi - \chi)} \, d \chi\,.}
where $\rho_i(\chi)$ is the linear charge density on the ring in the flat
$\IR^4$ metric.  If one expands $\rho_i(\psi)$ into a Fourier series:
\eqn\rhoFS{
\rho_i(\psi) ~=~ a_0^i + \sum_{n=1}^ \infty \big(\, a_n^i \, \cos(n\,\psi) ~+~
b_n^i \, \sin(n\,\psi)\,\big) \,,}
then the integrals in \lambdaform\ become elementary contour integrals and one finds
\eqn\lambdaseries{\lambda_i (z,r,\psi) ~=~{\pi \over R}\, (x-y) \,  S_i(y,\psi) \,,}
where $y$ is the coordinate defined in \xycoords, and
\eqn\Sseries{ S_i(y,\psi) ~\equiv~ a_0^i ~+~
\sum_{n=1}^ \infty \, \Big({y+1 \over y-1} \Big)^{n/2} \,\big(\, a_n^i \, \cos(n\,\psi) ~+~
b_n^i \, \sin(n\,\psi)\,\big) \,,}
Note that as $y \to - \infty$, $S_i(y,\psi) \to \rho_i(\psi)$.  That is, as one
approaches the ring, the function $S_i$ limits to the charge density.

Thus we take:
\eqn\Zsoln{Z_i  ~=~ 1 ~+~{\pi \over R }\, (x-y)\, S_i(y,\psi) ~-~ 
{4 q_j \, q_k \over R^2} \, (x^2 - y^2)    \,,}
where $i,j,k$ are all distinct, and the term proportional to 
$q_j \, q_k$ is a consequence of the source term in \Zeqn.

One of the very useful properties of the coordinates $(x,y,\psi, \phi)$
is that the Laplacian is separable.  More precisely, one can separate
variables in the function $F(x,y,\psi, \phi)$ if one acts on $(x-y) \, F$
with the Laplacian:
\eqn\sepLap{\eqalign{ & {R^2 \over (x-y)^3}   \, \nabla^2 [\, (x-y)\, F \,] ~=~  \cr
 & \quad \partial_x \big( (1 - x^2 ) \,  \partial_x \, F \big) ~+~ 
\partial_y \big( (y^2 -1) \,  \partial_y \, F \big) ~+~ (y^2 -1)^{-1}\,
\partial_\psi^2 \, F   ~+~ (1 - x^2 )^{-1}\,
\partial_\phi^2 \, F  \,.}}
Thus it is relatively easily to find general solutions for the $Z_i$ in terms
of orthogonal eigenfunctions.  In particular, it is elementary to show directly from
\Zeqn\ that  $S_i$  must satisfy:
\eqn\baseDE{ (y^2 -1) \,  \partial_y \big( (y^2 -1) \,  \partial_y \, S \big)
~+~ \partial_\psi^2 \,  S ~=~ 0 \,,}
away from the ring.

It is now convenient to separate out the constant modes and normalize them to
the conventions of \BenaDE, and so we set $a_0^i = {Q_i \over \pi}$. 
It is also useful to introduce the functions:
\eqn\Omseries{ \Omega_i(y,\psi) ~\equiv~  
\sum_{n=1}^ \infty \, \Big({y+1 \over y-1} \Big)^{n/2} \,{1 \over n}\, 
\big(\, a_n^i \, \sin(n\,\psi) ~-~ b_n^i \, \cos(n\,\psi)\,\big) \,.}
These functions are the indefinite integrals with respect to $\psi$ of the
oscillatory part of $S_i$.  They also satisfy the differential equation \baseDE.
One therefore has:
\eqn\Zsoln{Z_i  ~=~ 1 ~+~{Q_i \over R }\, (x-y)  ~-~ {4 q_j \, q_k \over R^2} \, 
(x^2 - y^2)   ~+~  {\pi \over R }\, (x-y)\,  \partial_\psi \Omega_i(y,\psi) \,.}
Following \BenaDE\ we also define:
\eqn\ABCdefn{A ~\equiv~  2 (q_1+q_2+q_3)\,, \quad B ~\equiv~ 
{2\over R} (Q_1\, q_1 + Q_2\, q_2 + 
Q_3\, q_3)\,, \quad  C~\equiv~ -  {24 q_1\, q_2\, q_3 \over R^2}\,.}

The final step is to solve \fundeqnsc.  To this end we make an Ansatz:
\eqn\kAnsatz{ k ~=~ k_0 \, dy ~+~ k_1 \, d\psi ~+~ k_2 \, d \phi \,.}
One should note that there is a gauge freedom in the definition 
of $k$ since a change of coordinate $t \to t + g$  generates a 
gauge transformation $k \to k + dg$.  The absence of a $dx$ term in
\kAnsatz\ may be viewed as a gauge choice.  One then obtains 
the system of differential equations:
\eqn\ksystem{\eqalign{ &\partial_y k_1 ~-~ \partial_\psi k_0~-~ \partial_x 
k_2   ~=~ - \big[ A ~+~B\,(x-y) ~+~ C\,(x^2 - y^2) ~+~ \coeff{2\,\pi}{R}\,
(x-y)\, \partial_\psi \Omega(y,\psi) \,\big] \,,  \cr  
&\qquad\qquad  \qquad \qquad (1-x^2)(y^2 -1) \, \partial_x k_0 ~-~ 
\partial_\psi k_2 ~-~   \partial_\phi k_1   ~=~ 0 \,, \cr
& \qquad \qquad \qquad \qquad (y^2 -1)\,(\partial_y k_2 - \partial_\phi k_0) ~+~ 
(1-x^2)\,  \partial_x  k_1  ~=~ 0\,,}}
where 
\eqn\Omdefn{\Omega ~\equiv ~ \sum_i q_i\, \Omega_i\,.}

It is easy to see that a solution to this system is given by:
\eqn\ksols{\eqalign{  k_0 ~=~   & -\coeff{\pi}{R}\, \big[ x\,(y^2-1)^{-1}\, 
\partial^2_\psi \Omega ~+~ 2\, y\, \Omega\, \big]\, \,,  \cr  
 k_1 ~=~  & (y^2-1)\,\big(\, \coeff{1}{3}\,C \,(x+y) ~+~
\coeff{1}{2} \,B \big) ~-~ A\, (y+1)~+~\coeff{\pi}{R} \, x\, (y^2 -1) \,
\partial_\psi \partial_y \Omega \,,  \cr   
k_2 ~=~ &  (x^2-1)\,\big(\, \coeff{1}{3}\,C \,(x+y) ~+~
\coeff{1}{2} \,B \big)  ~-~\coeff{\pi}{R} \, (1- x^2 ) \,
\partial_\psi \Omega   \,.}}
To show that this solves \ksystem\ one merely needs to use the
fact that $\Omega$ satisfies \baseDE.  If $\Omega = 0$ then this is simply 
the solution of \refs{\BenaDE,\ElvangDS,\GauntlettQY}. 
 
 In principle one can add an arbitrary homogeneous solution
 of \ksystem\ to \ksols.  However, this homogeneous solution is then fixed by 
 requiring that there are no CTC's near the ring.  This was used in \BenaDE\  
 to fix the polynomial behavior of the  $k_j$ as functions of $y$.  The ring is located
 at $y = -\infty$, and one can readily verify that $\Omega$,   $\partial_\psi^m 
 \Omega $ and $(y^2 -1) \, \partial_\psi \partial_y \Omega$ are
 all finite as $y \to - \infty$, and so it is unlikely that these new terms will generate 
 any new CTC's near the ring, and so no further additions of homogeneous
 solutions should be necessary. We will see this more explicitly below.

It turns out that there is a more convenient gauge choice for $k$, one in which 
the $y$-component of $k$ is zero:
\eqn\newkgauge{k ~=~ \hat k_0 \, dx ~+~ \hat  k_1 \, d\psi ~+~  \hat  k_2 \, d \phi \,.}
This may be achieved by taking $k \to k+ dg$. where
\eqn\gaugetrf{g~=~ -\coeff{\pi}{R} \, x\,(y^2 -1)\, \partial_y \Omega ~+~  
\coeff{2\, \pi}{R}\, \int y\, \Omega \, dy  \,.}
One then finds:
\eqn\khatsols{\eqalign{ \hat k_0 ~=~   & -\coeff{\pi}{R} \,(y^2-1) \, 
\partial_y\Omega   \,,  \cr  
\hat k_1 ~=~  & (y^2-1)\,\big(\, \coeff{1}{3}\,C \,(x+y) ~+~ \coeff{1}{2} \,B \big) ~-~ 
A\, (y+1)~+~\coeff{2\, \pi}{R} \, \int y\,  \partial_\psi \Omega \, dy  \,,  \cr   
\hat k_2 ~=~ &  (x^2-1)\,\big(\, \coeff{1}{3}\,C \,(x+y) ~+~
\coeff{1}{2} \,B \big)  ~-~\coeff{\pi}{R} \, (1- x^2 ) \,
\partial_\psi \Omega   \,.}}
We also need to fix the function of $\psi$
that appears as a ``constant of integration'' from $\int y \Omega  dy $ 
in \gaugetrf.  This is elementary:  One must not have any strings, and
so this indefinite integral must vanish at spatial infinity, or as
$y \to -1$.  It then follows from \Omseries\ that this indefinite 
integral has a leading behavior:
\eqn\intOm{\int y\, \Omega \, dy  ~\sim~  (y+1)^{3/2}\quad {\rm as} \quad  y \to -1\,. }

Indeed, to verify that there are no strings within our solution one must 
check  that $\hat k_1$ vanishes as $z \to 0$, or $y  \to -1$, and
$\hat k_2$ vanishes as $r \to 0$, or $x  \to -1$.  This is evident from
\khatsols\ and \intOm.

We now have the complete solution with arbitrary charge densities.  
 
 \subsec{The  near-ring  limit}

To examine how the metric behaves near the ring we want to take the limit
as $y \to - \infty$.   This structure of the metric is simpler to disentangle if we use
the rotation vector \khatsols, instead of \ksols.    Observe that the gauge 
transformation \gaugetrf\  has added to $k$ a term that diverges as $y \to - \infty$.  
Indeed, the leading  divergence is simply:
\eqn\leaddiv{\coeff{\pi}{R} \, y^2 \,\Big[\, \lim_{y\to - \infty} \,\partial_\psi 
\Omega(y,\psi)\,\Big]  ~\sim~ \coeff{\pi}{R}  \, y^2 \, \sum_i q_i \hat \rho_i(\psi) \,.}
where 
\eqn\hatrhodefn{\hat \rho_i(\psi) ~\equiv~ \rho_i(\psi) - a_0^i}
is the oscillatory part of the charge density.  Recall that the linear charge 
densities on the ring are ${1 \over \pi} \widehat Q_i(\psi)$, where
\eqn\Qhats{ \widehat Q_i(\psi) ~\equiv~ Q_i ~+~ \pi \, \hat \rho_i(\psi) \,.}
Therefore by making the replacement
\eqn\Qrepl{ Q_i \to \widehat Q_i(\psi) \,, }
we can incorporate the leading divergence, \leaddiv, in $k$
into  the coefficient $B$ defined in \ABCdefn, and thus reduce the problem
almost to that of \BenaDE.

We now consider the three-dimensional spatial part of the metric:
\eqn\threemet{\eqalign{
ds^2_3 ~=~ &    - (Z_1\, Z_2\, Z_3)^{-2/3} \, k^2  \cr & ~+~ 
{R^2 \over (x-y)^2}\,   (Z_1\, Z_2\, Z_3)^{1/3}  \,  
\Big(  (y^2-1)\, d \psi^2  +{dx^2 \over 1-x^2} + (1-x^2) \, d \phi^2  \Big)  }}
as we approach the ring.  Since $Z_i \sim y^2$ and $ k \sim y^3$ as $y \to - \infty$ 
this metric potentially diverges as $y^2$.  However one finds that the terms 
that are quadratic and linear in $y$ exactly cancel, leaving a finite part:
\eqn\asympmet{ 
ds^2_3 ~=~    \Big({C^2 \over 9 \,R^2}\Big)^{1/3}\, \Big[ \Big({9 \over C^2}\Big) \,
\widehat M(\psi) \, d \psi^2 ~+~ R^2 \, \left(d \theta^2 + \sin^2 \theta \, 
(d\phi + d \psi)^2\right)\Big] \,,}
where we have set $x = - \cos \theta$.  The function, $\widehat M(\psi)$,
is almost the obvious generalization of the parameter $M$ in \BenaDE:
%\foot{There was a typographical error in the first version of \BenaDE}:
%
\eqn\Mhatdefn{  \widehat M ~\equiv~  (2 q_1 q_2  \whQ_1  \whQ_2 + 2 q_1 q_3  
\whQ_1  \whQ_3 + 2 q_2 q_3    \whQ_2  \whQ_3 - q_1^2  \whQ_1^2 - q_2^2 
\whQ_2^2 - q_3^2  \whQ_3^2) +  \coeff{1}{3}\, C \,R^2 (A - 2\, \alpha(\psi)) \,,}
where $\whQ_i$ is defined by \Qhats, and $\alpha(\psi)$ is defined via:
\eqn\alphadefn{ \partial_\psi \Omega ~=~ \Big(\sum_i  q_i \hat \rho_i(\psi) 
 \Big) ~+~   \coeff{R}{\pi}  \,    \alpha(\psi) \, y^{-1} ~+~
{\cal O}(y^{-2}) \quad {\rm as} \quad y \to -\infty \,.}
In terms of the Fourier series we have 
\eqn\alphaseries{  \alpha(\psi) ~=~  \sum_{i=1}^3 \, q_i \,  \alpha_i(\psi)
~=~ {\pi \over R}\, \sum_{i=1}^3 \, q_i \,   \sum_{n=1}^ \infty n\,  \big(\, a_n^i \, 
\cos(n\,\psi) ~+~ b_n^i \, \sin(n\,\psi)\,\big) \,.}

We therefore see that, with the exception of the $\alpha(\psi)$ term, the
form of the near-ring metric is exactly that of \BenaDE, but with the replacement
\Qrepl.  This makes perfect physical sense:  The functions $\whQ_i(\psi)$
are the local charge densities on the ring, and so if one approaches the 
ring closely then the ring should look like the infinitely long black tube \BenaWV\
with charge density given by the local values.

The only unexpected modification is the appearance of the $\alpha(\psi)$ term
as a shift in $A  \equiv   2 (q_1+q_2+q_3)$.  The sum of the $q_j$ is 
essentially the angular momentum of the black ring, and so $\alpha(\psi)$ appears
to be  a local shift in the angular momentum.   As we will see below, none
of the fluctuations like $\alpha(\psi)$ are visible from infinity.   Thus, this 
shift must be related to the angular momentum balance of the ring locally
on the ring surface.   One should also note that while the ring is a perfect circle 
in the original $\IR^4$ base, it is no longer  circular in the complete
metric.  From \asympmet\ one can see that its radius of curvature is varying:
\eqn\varyingR{ \mu(\psi) ~=~ \Big|{3 \over C}\Big|  \, \sqrt{\widehat M(\psi) } \,.}
Therefore one should expect some fluctuation in the angular momentum 
contribution to the horizon area.  
% It would be interesting to see if one can
%see this more directly through some Born-Infeld analysis of the brane
%dynamics.

%%%%%%%%%%%%%%%%%%%%%%%%%%%%%%%
\goodbreak\midinsert
\vskip .2cm
\hskip .2cm
\centerline{ {\epsfxsize 5.2in\epsfbox{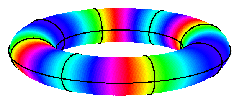}}}
\vskip -.25cm
\leftskip 2pc
\rightskip 2pc\noindent{\ninepoint\sl \baselineskip=8pt
{\bf Fig.~1}:  The horizon of the black ring.  The cross-section is
really an $S^2$ of fixed radius, $R$, but is depicted here as an $S^1$. 
The variation of  $g_{\psi \psi}$  along the  $\psi$-direction in 
the horizon metric \asympmet\ is depicted as a change of color.
}
\endinsert
%%%%%%%%%%%%%%%%%%%%%%%%%%%%%%%%%%%

\subsec{Asymptotic charges}

The region far from the ring corresponds to $x = -1, y=-1$.  Indeed, in this
limit:
\eqn\asympxy{x ~\sim~ -1 ~+~ {2 \, R^2 \,r^2 \over
(z^2 + r^2)^2 }\,, \qquad y ~\sim~ -1 ~-~ {2 \, R^2 \,z^2 \over
(z^2 + r^2)^2 } \,.}
Therefore, the asymptotic charges of the solution come from the terms in the 
electric potentials that fall-off as $(y+1)$.  However, 
 the $(x-y) \partial_\psi \Omega$ terms in
\Zsoln\ fall-off faster than that, and therefore the electric charge measured at
infinity is exactly that of the $U(1) \times U(1)$ invariant ring 
\refs{\BenaDE,\ElvangDS,\GauntlettQY}:
\eqn\Qinfinity{ Q^{\infty}_i ~\equiv~ Q_i~+~ 4 \, \sum_{j,k}\, |\epsilon^{ijk}| \, q_j \, q_k \,}
One can convert  these charges to number densities of branes using:
\eqn\QtoN{
Q_i ~=~  {\bar N_i l_p^6 \over 2 L^4 R}\,,\qquad q_i={n_i l_p^3 \over 4 L^2}\,,}
where $L$ is the length of the two-tori. The asymptotic charges, 
$N_i$, of the solution are then the sum of the charges 
on the black ring $\bar N_i$, and the charges dissolved in fluxes:
\eqn\asympN{
N_1 = \bar N_1 + n_2 n_3 ~,~~~ N_2 = \bar N_2 + n_1 n_3 ~,~~~ 
N_3 = \bar N_3 + n_1 n_2  \,.}

The angular momenta  also behave in exactly the same way:
They can be read off from the $dt \, d\psi$ and $dt \, d\phi$ terms in the metric that 
vanish as  $(1+x)^a  (y+1)^{(1-a)}$, for some $a$, or inversely as the square of the 
distance.  From \khatsols\ and \intOm\ one sees that the terms involving $\Omega$
do not contribute; hence the angular momenta are also the same as for the 
uniformly charged ring:
\eqn\angmom{
J_1 ~=~  J^T ~+~ {1\over 2}
\Big( \sum_{i=1}^3\, {n_i N_i} ~-~ n_1 n_2 n_3 \Big)\,, \qquad 
J_2 ~=~   - {1\over 2}
\Big( \sum_{i=1}^3\, {n_i N_i} ~-~ n_1 n_2 n_3 \Big) \,,}
where $J^T$ is the angular momentum carried by the ring:
\eqn\angmomring{J^T  = {R^2 L^4 \over l_p^6}\,(n_1+n_2+n_3)\,.}

Thus the varying charge densities are undetectable in the asymptotic
charges, and so the fluctuating charge densities, $\hat \rho_i(\psi)$, 
truly represent an infinite amount of ``hair'' on the black ring.  

%%%%%%%%%%%%%%%%%%%%%%%%%%%%%%%%%%
\newsec{Some properties of the solutions}
%%%%%%%%%%%%%%%%%%%%%%%%%%%%%%%%%%

We have exhibited a family of solutions with three arbitrary fluctuating
charge densities, $\hat \rho_i(\psi)$, and the asymptotic charges of
the solution are completely independent of these fluctuating densities.
On the other hand, the horizon geometry, and its area, do depend non-trivially
upon these functions.  
In particular, the function $\widehat M(\psi)$  is a quadratic functional of
these densities.  Indeed, the  $\whQ_i$ (which appear quadratically in \Mhatdefn)  
are linear functions of the $\hat \rho_i$, and the function
$\alpha(\psi)$ is also a linear functional of $\hat \rho_i(\psi)$, as can be seen 
from \alphaseries.

Supertubes are, of course, obtained by setting $\widehat M(\psi)$ identically
equal to zero.  Since this involves one functional constraint on three arbitrary
functions, our solutions will contain a family of supertubes parametrized by 
two independent 
functions.  Before examining this more closely, it is interesting
to broaden the issue a little, and discuss families of solutions for which 
$\widehat M(\psi)$ is actually a constant, and not necessarily zero.   
These solutions are physically very interesting because, even though the 
complete solution is $\psi$-dependent and breaks  $\psi$-translation 
symmetry, the {\it horizon} does have $\psi$-translation 
symmetry if $\widehat M(\psi)$ is constant. Thus the geometry near the horizon 
is identical to that of the round ring, and the solutions  provide a 
very graphic example of black hole hair, as they have two functions-worth of hair
that lives on a completely round ring.

As an example, consider charge density modes that only have a finite number of 
Fourier terms - that is, the series \rhoFS\ ends at $n= K$.  Then one can simply 
substitute  \rhoFS\  and \alphaseries\ into  $\widehat M(\psi)$.  The result is 
then a Fourier series in $\sin(n \psi)$ and $\cos(n \psi)$ for $n \le 2K$.  Setting
all but the constant term to zero will impose $4 K$ constraints on the 
$6 K$ variables $a_n^i$, $b_n^i$, $ n=1, \dots,K$, $i=1,2,3$.  Thus we
get $2 K$ free parameters in general.   One should also note that 
in simplifying $\widehat M(\psi)$, one will use the identities 
$\cos^2(n \psi) = {1 \over 2} (1 + \cos(2 n \psi))$, 
$\sin^2(n \psi) = {1 \over 2} (1- \cos(2 n \psi))$, and this will shift the
constant terms in $\widehat M(\psi)$ by ${1 \over 2} (a_n^i)^2$ and 
${1 \over 2} (b_n^i)^2$.  Thus the fluctuating modes {\it will} contribute to the
horizon area.

It is elementary to implement the foregoing analysis for a single Fourier mode.  
Take $a_n^i = {R \over \pi} c_i$ and $b_n^i = {R \over \pi} d_i$ for
some $n$, and set all the other modes to zero.  Then
\eqn\whQform{
\whQ_i ~=~ Q_i ~+~ R\,(c_i\, \cos(n \psi) + d_i \, \sin(n \psi)) \,.}
and
\eqn\alphaform{\alpha(\psi) ~=~ \sum_{i=1}^3 \, q_i \, n
\,(c_i\, \cos(n \psi) + d_i \, \sin(n \psi)) \,.}
Let $\vec c$ and $\vec d$  be the vectors whose components are
$c_i$ and $d_i$, and define:
\eqn\qvecs{\vec Q ~=~ (Q_1\,, Q_2\,, Q_3)\,, \qquad \vec q ~=~ (q_1\,, q_2\,, q_3)\,, 
\qquad \vec q_c ~=~ \coeff{1}{R}\, (q_2 q_3\,, q_1 q_3\,, q_1 q_2)  \,.}
Introduce the matrix:
\eqn\Mmat{{\cal M} ~\equiv~  \left( \matrix{-q_1^2 & q_1\, q_2 &q_1\, q_3 \cr 
q_1 \, q_2 & -q_2^2 & q_2\, q_3 \cr  q_1 \, q_3  & q_2 \, q_3 & -q_3^2  }  \right)  \,,}
then the condition that $\widehat M(\psi)$ is independent of $\psi$ yields the
equations:
\eqn\Mmat{ \eqalign{
(\vec Q + 8\, n\, \vec q_c)^T \cdot {\cal M} \cdot \vec c ~=~& 0 \,, \qquad    
(\vec Q + 8\, n\, \vec q_c)^T \cdot {\cal M} \cdot \vec d ~=~0 \,, \cr
(\vec c)^T \cdot {\cal M} \cdot \vec d ~=~ & 0 \,, \qquad (\vec c)^T \cdot {\cal M} \cdot 
\vec c ~=~ (\vec d)^T \cdot {\cal M} \cdot \vec d  \,.}}
One also finds that the value of $\widehat M$ is shifted from its value when
$\hat \rho_i =0$ by an amount:
\eqn\deltaM{  
\Delta  \widehat M ~=~  \coeff{1}{2}\, R^2 \, \Big[\, (\vec c)^T \cdot {\cal M} \cdot \vec c  
~+~(\vec d)^T \cdot {\cal M} \cdot \vec d \,  \Big] \,.} 
Therefore, in the (indefinite) metric defined by ${\cal M}$, the vectors 
$(\vec Q + 8\, n\, \vec q_c)$, $\vec c$ and $\vec d$ must all be mutually orthogonal, 
and $\vec c$ and $\vec d$ must have the same norm.  There are thus two parameters:
The norm of $\vec c$ and the freedom to make a rotation in the $(\vec c, \vec d)$ plane.
The shift in the horizon parameter $\widehat M$ is proportional to the norm-squared
of $\vec c$.  

If one has $Q_i = Q$ and $q_i =q$, $i =1,2,3$, then, in the standard inner
product in $\IR^3$ the vectors $\vec c$,  $\vec d$ and $(1,1,1)$ must be 
mutually orthogonal, with $\vec c \cdot \vec c  = \vec d \cdot \vec d$, and then
\eqn\whMsimp{\widehat M ~=~ 3 \, q^2 \, (Q^2 - 16\, q^2) ~-~ 
q^2\, R^2 \, \big[\,  \vec c  \cdot  \vec c   ~+~\vec d  \cdot  \vec d   \,  \Big] \,.}
%

%%%%%%%%%%%%%%%%%%%%%%%%%%%%%%%%%%
\newsec{A conformal field theory on the ring?}
%%%%%%%%%%%%%%%%%%%%%%%%%%%%%%%%%%

The calculation outlined in the last section actually amounts to solving the classical
Virasoro constraints for two-dimensional  scalar  fields whose target space
is  a $(2+1)$-dimensional, Lorentzian space.  This  suggests that one might 
understand the ``hair'' defined by the charged modes in terms of a conformal field theory.

Introduce three scalar fields:
\eqn\scalarfields{
X^j (w, \bar w) ~=~ \coeff{1}{2} \, \alpha_0^j  \,  \log(w\, \bar w)~+~ \sum_{n\ne 0} 
\,\coeff{1}{n}\, \big[\, \alpha_n^j \, w^n ~+~ \tilde \alpha_n^j \, \bar w^n \,\big]\,,}
where, for the moment, the sum over $n$ runs from $-\infty$ to $\infty$.
Let $w = e^{- \tau + i \psi}$, and observe that the $X^j (w, \bar w)$ satisfy
the harmonic equation:
\eqn\harmeqn{( \partial_\tau^2 ~+~ \partial_\psi^2) \, X^j  ~=~ 0 \,.}
To relate this to the results of the previous section, note that if one makes
the change of variable:
\eqn\CofV{ y ~=~ - {\cosh \tau \over \sinh \tau} \,,}
then equation \baseDE\ becomes exactly the harmonic equation \harmeqn.
We now rescale the functions $S_j$ of the previous section, and 
set:
\eqn\SandX{\pi \, q_j \, S_j ~=~ \partial_\tau \, X^j \,,}
with no sum on $j$.  In making this identification we must set
$ \alpha_n^j = \tilde \alpha_n^j = 0$ for $n <0$ and impose
the reality condition $\tilde \alpha_n^j  = (\alpha_n^j)^*$.
Setting the negative modes to zero is required in order to make
the solution regular at $\tau = \infty$, or $y =-1$, and in particular, this
means that the solution is  regular at spatial infinity in the original metric \firstRmet.

Introduce the matrix:
\eqn\Pmat{{\cal P} ~\equiv~  \left( \matrix{1  & -1 & -1 \cr 
-1 & 1 & -1  \cr  -1  &  -1  & 1  }  \right)  \,,}
and  define: 
\eqn\Tdefn{T ~\equiv~  - ( \partial_\tau \,  \vec  X)^T  \cdot {\cal P} \cdot
( \partial_\tau \,  \vec  X)  ~-~  {16\, q_1\, q_2\, q_3 \over R} \, \bigg( \sum_{j=1}^3\, 
( \partial_\tau^2 \,   X^j) \bigg)~-~ 16\,  q_1\, q_2\, q_3\,(q_1+  q_2+ q_3) \,.}
Then one can easily check that:
\eqn\magic{\widehat M(\psi) ~=~ \lim_{\tau \to 0} \,  T \,.}

Observe that ${\cal P}$ has eigenvalues $-1, +1, +1$ and so defines a Lorentzian 
metric  on $\IR^3$.  Moreover, $T$ represents the kinetic part of an (indefinite) 
Hamiltonian for a scalar field theory  with a ``charge at infinity.''  That is,  $T$ contains 
a kinetic term for the scalar fields, but does not contain the ``elastic terms,''  
$( \partial_\sigma X^j )^2$.   However, if one only  considers solutions with a round horizon, 
and hence constant $\widehat M(\psi)$, then in terms of $T$, this condition amounts 
to the Virasoro constraints:
\eqn\Vir{L_n ~\equiv~  {1 \over 2\,\pi}\, \int_0^{2\, \pi} \, e^{i   n  \psi} \, 
T(\tau , \psi)|_{\tau=0}  \,  d \psi ~=~ 0 \,,  \qquad n >0 \,.}

This Virasoro condition reduces the degrees of freedom by one
bosonic function. Note that if this were a string theory, then $T$ would
contain energy terms of the form, $( \partial_\sigma X^j )^2$, and the 
theory would possess  conformal/reparametrization invariance. The latter
would reduce the degrees of freedom by one additional function, leaving
only one transverse set of string modes\foot{ This is typically used
in light-cone gauge to set $X^+ \sim p^+\, \tau$.}.  Here we do not appear to 
have this additional harmonic reparametrization invariance,  
$\tau \to \tilde \tau(\tau, \psi)$  and  $\psi \to \tilde \psi(\tau, \psi)$, because
the other physical fields, like the metric \newRmet, are not
invariant under such reparametrizations.  Thus, one should probably
think of this as free bosons on a space with indefinite signature,
and there are still two freely choosable arbitrary functions in the solution after 
one solves \Vir.   

For $\widehat M(\psi)=0 $, the free bosons described above will encode  a lot of 
zero entropy configurations that one might use to explain part of the entropy of 
black holes and black  rings. One of the main problems with counting three-charge 
supergravity configurations  is quantizing them.  If one can reinterpret the foregoing
classical free bosons in terms of a conformal field theory  then this may well 
provide a natural way to quantize  geometries, similar to that in \LinNB.

%%%%%%%%%%%%%%%%%%%%%%%%%%%%%%%%%%
\newsec{Final comments}
%%%%%%%%%%%%%%%%%%%%%%%%%%%%%%%%%%

We have constructed a huge family of supersymmetric black ring solutions with 
arbitrarily varying charge densities around the ring.  These solutions do not appear to
have any CTC's, and indeed bear out the intuition that if one approaches
the ring closely then it will look like an infinite black cylinder whose charge
density is set by its local value.  In particular, we found that the near-ring metric could
essentially be obtained from that of \refs{\BenaDE,\ElvangDS,\GauntlettQY} by replacing 
the constant 
charge  densities by the varying density functions, $\whQ_i(\psi)$.  The only correction
to this prescription involves an angularly dependent shift in the contribution
to the horizon area coming from the angular  momentum of the ring.    
While the ring has  a lot of structure coming from the fluctuating
charge densities, $\hat \rho_i(\psi)$, none of this structure contributes to
the asymptotic charges.  The charges of our new black rings are exactly those
of the $U(1) \times U(1)$ invariant black ring, and so the functions  $\hat \rho_i(\psi)$ 
do indeed represent non-trivial ``hair.''

Even if the shape of the dipole branes forming the ring is round in the 
flat $\IR^4$ base, the varying charge
densities feed back into the metric, so that the radius of curvature of the ring 
fluctuates. Indeed, the ring radius, $\mu(\psi) \sim \sqrt{\widehat M(\psi)}$, is quadratic 
in the fluctuating densities, $\whQ_i(\psi)$.  Thus a fluctuation   
in $\whQ_i(\psi)$ that has $n$ nodes could give rise to a fluctuation in the
ring radius that has $2 n$ nodes (depending upon whether the quadratic
on linear terms dominate). By tuning only one parameter we can obtain 
solutions that interpolate between black ring and supertube as one goes 
along the ring. While our analysis does not detect any obvious problem
with such solutions (apart from the well-known null orbifold), one might 
reasonably expect a more detailed analysis of the geometry to reveal
some singular behavior near the transition points. 

Probably the most physically relevant subset of our solutions has constant 
$\widehat M(\psi)$,  and is parametrized by two arbitrary functions. 
If $\widehat M=0 $ these solutions  are three-charge supertubes.
If $\widehat M > 0$ then these solutions describe black rings whose 
near-horizon geometry is the same as that of the $U(1) \times U(1)$ invariant ring,
which means that at least  the rings with  constant $\widehat M$ should
have regular horizon  structure.  We have also found that these solutions can be 
described by a set of three free bosons satisfying a Virasoro constraint and
with a target space of signature (2,1). 

Solutions with constant $\widehat M $ are also a good starting point for constructing 
non-BPS three-charge black rings (generalizing the  rings found in
\refs{\EmparanWN\ElvangYY\ElvangMJ{--}\EmparanWY}). 
A variable  $\widehat M$ non-extremal ring would  almost certainly be time dependent:  
The rotating lumps would  radiate energy away as gravitational waves, and the solution
would evolve to a rotating ring  in which he horizon was ``round.''
By the same token, one might also wonder if
non-BPS solutions with varying charge densities, but constant 
horizon radius,  would  be unstable through electromagnetic radiation,
and thus decay into states with uniform charge densities. This would be 
very interesting to pursue further. 
 
In \BenaDE\ it was shown that the most general  black-ring solution is given by 
seven arbitrary functions:  the four independent  profile functions, $\vec y(\psi)$,  and the 
three charge densities. It would be very  interesting to find this general solution 
explicitly, and to see if the configuration with a round horizon is similarly 
described by free bosons with a seven-dimensional target space. 
Understanding  this sigma model might be the key to quantizing the 
shape and density profiles that give  three-charge geometries.   
The entropy of our rings is an integral of a functional of the charge densities 
and can be freely varied by changing  these densities. This freedom persists in 
the two parameter family  of solutions that are $U(1) \times U(1)$ invariant 
near the horizon.  Even if the densities and the corresponding entropy are 
classically continuous,  at a quantum level one expects this continuity to be lost, 
as happens for  the  two charge supertube \MateosQS, where the discrete spacing 
between different shapes and density profiles accounts for the entropy of the tubes 
\refs{\PalmerGU,\BakKZ,\BakRJ}. Quantizing the space of supertubes in terms
of a CFT might open the door to computing the entropy 
in three-charge supertubes, and finding if this matches the black ring or 
black hole entropy.

Last, but not least, the $U(1) \times U(1)$ invariant rings have a very simple 
description in terms of the D1-D5 CFT \BenaTK. It would be nice to extend this 
description to the new black ring solutions we have constructed here, and to match 
their entropy to that of a sector of the D1-D5 CFT.

\bigskip
%%%%%%%%%%%%%%%%%%%%%%%%%%%%%%%%%%
\leftline{\bf Acknowledgments}
%%%%%%%%%%%%%%%%%%%%%%%%%%%%%%%%%%

We would like to thank Per Kraus, Gary Horowitz, 
Henriette Elvang and Roberto Emparan for helpful conversations and comments.
This work was supported in part by
funds provided by the DOE under grants DE-FG03-84ER-40168 and
DE-FG03-91ER-40662, and by the NSF under grants PHY00-99590 and
PHY01-40151. 

\vfill\eject
\listrefs
\vfill\eject
\end